\documentclass[runningheads]{llncs}
\usepackage{graphicx}
\usepackage{amsmath,graphicx}
\usepackage{tikz}
\usetikzlibrary{calc}
\usepackage{comment}
\usepackage{svg}
\usepackage{xcolor}
\usepackage{enumitem}
\usepackage[tableposition=top]{caption}

\usepackage[hidelinks]{hyperref}

\usepackage{caption}

\begin{document}
\title{Auto-segmentation of Hip Joints using MultiPlanar UNet with Transfer learning}
\titlerunning{Hip Joints AutoSeg}
%

\author{Peidi Xu\inst{1} \and
Faezeh Moshfeghifar\inst{1} \and
Torkan Gholamalizadeh\inst{1, 3}
\and Michael Bachmann Nielsen\inst{2} \and Kenny Erleben\inst{1} \and Sune Darkner\inst{1}
}

\authorrunning{P. Xu et al.}

\institute{Department of Computer Science,  University of Copenhagen, Denmark \email{peidi@di.ku.dk}\\
\and
Department of Diagnostic Radiology, Copenhagen University Hospital, Denmark \and 
3Shape A/S, Copenhagen, Denmark
}

\maketitle              
\begin{abstract}
Accurate geometry representation is essential in developing finite element models.
Although generally good, deep-learning
segmentation 
approaches with only few data
have difficulties in accurately segmenting fine features, e.g., gaps and thin structures.
Subsequently, segmented geometries need
labor-intensive manual modifications to reach a quality where they can be used 
for simulation purposes.
We propose a strategy that uses transfer learning to reuse datasets with poor segmentation combined with an interactive learning step where fine-tuning of the data results in anatomically accurate segmentations suitable for simulations.
We use a modified MultiPlanar UNet that is pre-trained using inferior hip joint segmentation
combined with a
dedicated loss function 
to learn the gap regions
and post-processing to correct tiny inaccuracies on symmetric classes due to rotational invariance.
We demonstrate this robust yet conceptually simple approach applied  
with clinically validated results on publicly available computed tomography scans of hip joints. 
Code and resulting 3D models are available at: \url{https://github.com/MICCAI2022-155/AuToSeg}

\end{abstract}
\section{Introduction}
Precise segmentation of medical images such as computed tomography (CT) scans, is widely used for generating finite element (FE) models of humans for patient-specific implants~\cite{chen2016image_guide}. A requirement in generating FE models is a proper geometrical representation of the anatomical structures~\cite{poelert2013patient}.
In our case, an \textit{accurate} segmentation of the hip joint (HJ) should essentially detail the shape and boundaries of the femur and hip bones and identify the inter-bone cavities. The segmented geometries should be closed, non-intersecting, and without spikes. As manual segmentation is labor-intensive and time-consuming~\cite{poelert2013patient}, automated segmentation tools are usually necessary to generate accurate FE models.

Convolutional Neural Networks with encoder-decoder structures are widely used for auto semantic segmentation, among which the most successful one is the UNet structure~\cite{unet_origin}.
The architecture uses skip connection on high-resolution feature maps in the encoding path to include more fine-grained information. Although more recent models are proposed on segmenting natural images, e.g., DeepLabV3+, UNet still provides some of the best segmentation results in medical images ~\cite{chen2018encoder_deeplabv3p} . Therefore, the variation of UNet, e.g., 3D UNet, is a straightforward way to segment 3D medical data like CT scans and has shown its state-of-the-art performance~\cite{cciccek2016_3dUnet_origin}. 
Applying 3D convolutions directly to large 3D images may overflow memory. Therefore, 3D models are usually trained on small patches, which results in a limited field of view and subsequent loss of global information. As an alternative with far less memory usage, the MultiPlanar UNet (MPUNet) model was proposed by Perslev et al.~\cite{perslev2019one_multiplanarUnet} which uses a 2D UNet to learn representative semantic information. 

Most studies on auto-segmentation of the HJs focus on designing more powerful neural networks that separate anatomical structures with little manual intervention~\cite{wang2018pelvis_shape_iterative,weston2020complete_show_data}. 
These studies focus primarily on the bone morphology and 
not on 
the inter-bone gaps. The consequence is that although they reach fairly high Dice scores, the segmentation results are anatomically inaccurate and are unsuitable for generating HJ 3D models. This limits the usability of the existing deep learning models for FE simulations~\cite{nishii2004three}.

\begin{figure}[htb]
\centering
\begin{minipage}[b]{0.9\linewidth}
  \centering
  \centerline{
  \includegraphics[width=\textwidth]{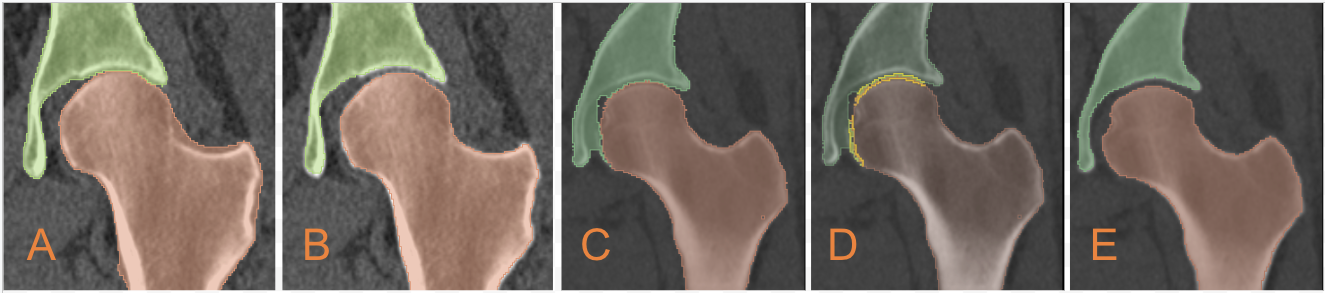}
  }

\end{minipage}
\caption{Illustration of gap generation: Inferior ground truth of a training image from public dataset (A) and results by fine-tuned model (B). Results on a test image with model trained only on public dataset (C) with  erroneous prediction detection (D) and fine-tuned (E)}
\label{fig:comparisons}
\end{figure}

We require the deep learning models to provide anatomically correct segmentation of the bones and the existing gap in the HJ as shown in Fig.~\ref{fig:comparisons}~\cite{wang2018pelvis_shape_iterative,weston2020complete_show_data}. Due to the limited number of accurate training data, we propose a deep learning-based strategy for enhancing publicly available poorly annotated scans using only a few accurately segmented data to learn an accurate model and in our case the gap regions in HJ. Besides using the idea of MultiPlanar, our backbone model is a standard UNet with batch normalization. Therefore, the proposed pipeline is both parameter and memory efficient.

To enforce the cartilage gap with few annotated data, we apply MPUNet with a dedicated loss function penetrated more on the gap regions combined with transfer learning and a post-processing step. Our framework uses an interactive learning pipeline involving pre-training MPUNet on a public dataset with inferior HJ segmentation to learn general semantic features of the bones~\cite{CTPEL}. The model is then fine-tuned using a few highly accurate segmentation to learn the correct labeling of the gaps. We show that our proposed approach allows the model to learn the gap and generate anatomically accurate segmentation, using the pre-trained model and only four accurate segmentations for fine-tuning. Our work is validated on a set of HJs from which we construct FE models and report the Dice with the manually corrected segments used for biomechanical models.

\section{Method}\label{sec:method}
Our strategy for accurate HJ segmentation with very few accurate training images relies on the following: (i) we use the idea of MPUNet that segments 3D medical images using 2D models while preserving as much spatial information as possible by segmenting different views of the data. (ii) we use a relatively simple yet powerful backbone model for performing the segmentation to avoid overfitting and memory issues. (iii) we pre-train the model using publicly available datasets with poor labels, which are then fine-tuned with a very small set of accurately annotated data. (iv)
we use a dedicated weighted distance loss to enforce the gap between the bones. (v) we introduce a post-processing step that solves the internal problem of MPUNet on images with symmetric features.

\noindent\textbf{\emph{Model}}: 
As a baseline model, we use the MPUNet proposed by Perslev et al.~\cite{perslev2019one_multiplanarUnet} to segment the 3D HJs using 2D UNet while preserving as much 3D information as possible by generating views from different perspectives.
During training, the model $f(x; \theta)$ takes a set 2D image slices of size $w \times h$, from different views, and outputs a probabilistic segmentation map $P \in R^{w \times h \times K}$ for K classes for each slice. Standard pixel-wise loss function is then applied for back-propagation. Our experiment uses a standard categorical cross-entropy loss augmented by the weighted distance map. We found no improvement using a class-wise weighted cross-entropy loss or the dice loss.
In the inference phase, we run 3D reconstruction in each view separately over the segmentation results on all the parallel slices to get the volume back. This results in a volume probability map of size $m \times w \times h \times K$ for each view. Unlike original MPUNet \cite{perslev2019one_multiplanarUnet} which suggests training another fusion model using validation data, we simply sum over the results ($P$) from different views followed by an argmax over last dimension to get the final label map. This strategy achieves good results on the validation data.

\noindent\textbf{\emph{Transfer Learning}}: The accurate segmentation and fast convergence rely partially on pre-training the model using publicly available datasets with poor labeling, which is subsequently fine-tuned with a small set of accurate data. We detail two modifications that differ from standard transfer learning settings. First, we also transfer the weight in the last softmax layer for a much faster convergence because we work on exactly the same classes as before.  Then, instead of freezing encoder and only fine-tuning decoder, it is necessary to explicitly learn encoder to detect the gap, as the gap must be encoded correctly first.

\noindent\textbf{\emph{Weighted Distance Map}}: For the model to be fined-tuned to learn the gap between the bones, we enforce a voxel-wise weight-map $w(x)$ to the loss function based on the distances to the border of the foreground classes. This strategy was initially suggested in the original UNet paper, which we employ in a modified version for 3D data \cite{unet_origin,perslev2019one_multiplanarUnet}. We define $w(x)$ as follows,
\begin{equation}
\label{eq:weight_map}
    w(x) = w_c(x) + w_0 \cdot e^{-\frac{(d_1(x)+d_2(x))^2 }{2\sigma^2}}
\end{equation}
where $d_1$ and $d_2$ denotes the distance to the border of the nearest foreground class and the second nearest foreground class respectively. 
We follow original UNet paper and set $w_0 = 10$ and $\sigma = 5$. $w_c: \Omega 	\rightarrow  R$ is used to balance the class frequencies, which we do not enforce, thus we set $w_c = 1$ for every c.

\begin{figure}[htb]
\centering
\begin{tikzpicture}

\node [above right, inner sep=0] (image) at (0,0) 
{ 
\includegraphics[width=0.9\textwidth]{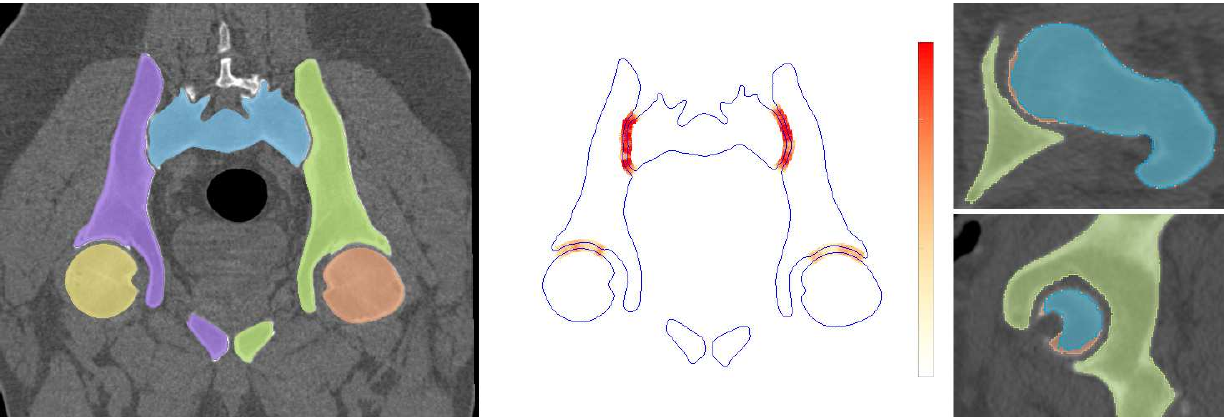}
};

\begin{scope}[
    x={($0.1*(image.south east)$)},
    y={($0.1*(image.north west)$)}]
 
\draw[ultra thin,black] (3.93,0.05) rectangle (7.73,9.88);
\node at (7.35,1.2){\tiny 1};
\node at (7.35,4){\tiny 2};
\node at (7.35,7.1){\tiny 3};
\node at (7.35,9){\tiny 3.7};
\end{scope}
\end{tikzpicture}
\caption{(left) A sample training slice of true labels overlaid on top of raw image. (middle) Corresponding weight map computed with Eq. \eqref{eq:weight_map} overlaid on top of label boundaries. (right) Results of training with weight map calculated over eroded labels(orange contour), which shows a smoother and more complete contour near the boundaries than the results trained without erosion (blue contour).  }
  \label{fig:weight_train}
\end{figure}
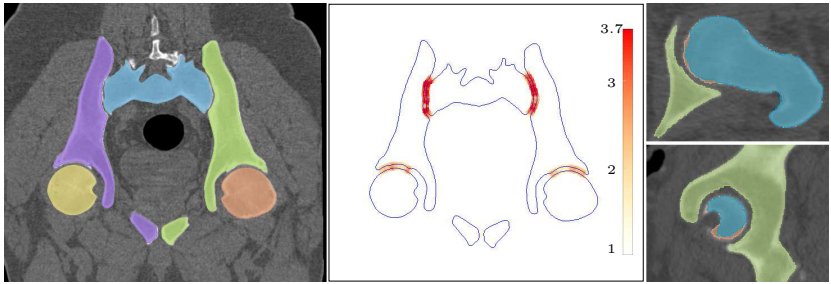

During fine-tuning, the corresponding slice of the 3D weight map is sampled together with the images and labels. We apply an element-wise multiplication of the weight map with the cross-entropy loss of predictions and labels on each pixel before reduction. Fig.~\ref{fig:weight_train} (left \& middle) shows an example training slice. Note that we do not plot the prediction since it consists of multi-class probabilities.

We also notice that the model is prone to overfitting to the gap, producing a broader gap than usual if we assign higher weights only to the gap regions in Eq.~\eqref{eq:weight_map}. Instead, we would like to assign more weight to the boundaries around the gap to avoid false negatives. This is accomplished by applying a mathematical erosion to the labels over a ball with a radius of 3 voxels before calculating the weight map, as demonstrated in Fig.~\ref{fig:weight_train} (middle). To compensate for the increased value of $d_1+d_2$ introduced by erosion, we double $w_0$ to $20$.

\noindent\textbf{\emph{Sampling Strategies}}: Sampling and interpolation are necessary to retrieve corresponding 2D slices from a 3D medical image viewed from a random orientation other than the standard RAS axes. We follow the idea in \cite{perslev2019one_multiplanarUnet} by sampling pixel with dimension $d \in \mathcal{Z}^+$ on isotropic grids within a sphere of diameter $m \in \mathcal{R}^+$ centered at the origin of the scanner coordinate system in the physical scanner space. We differ in that these two numbers are chosen as the 75 percentile across all axes and images during training but as maximum value during inference. This ensures both efficient training and complete predictions near the boundaries.

\begin{figure}[htb]
  \centering
  \centerline{
  \includegraphics[width=0.78\textwidth]{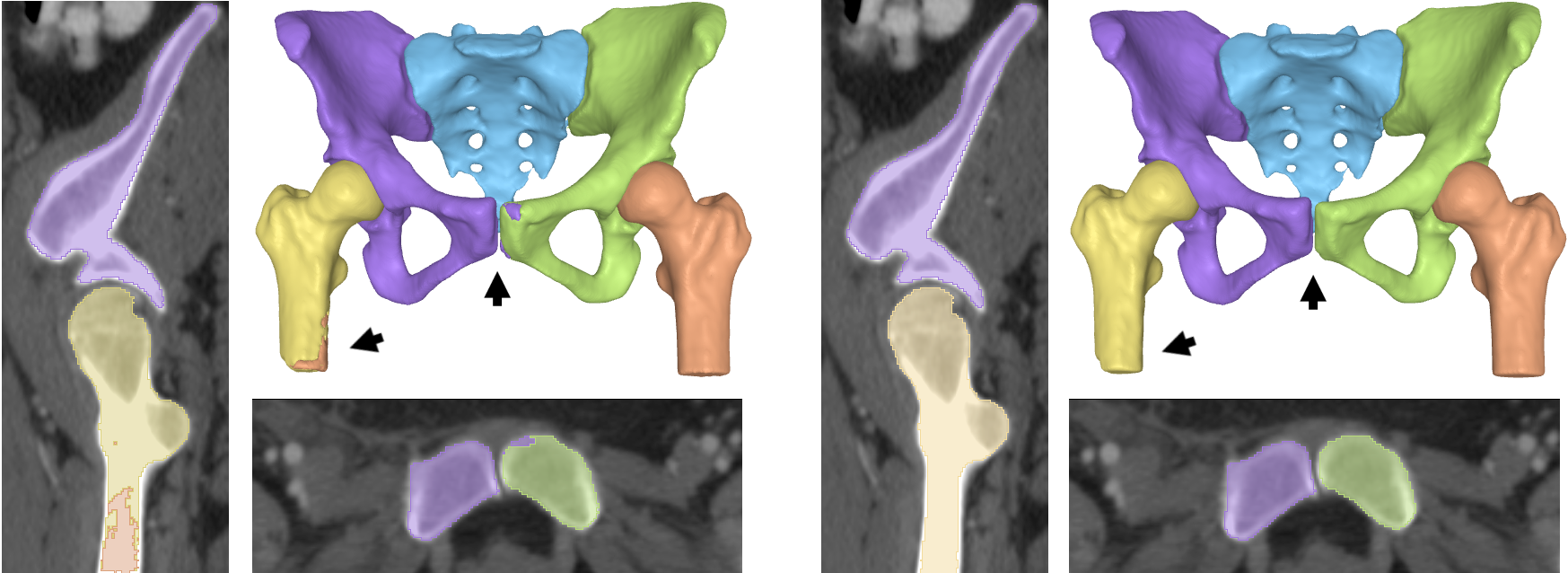}
  }

\caption{Segmentation (left) with post-processing (right) where falsely predicted symmetric groups are recovered.}
\label{fig:post_effect}
\end{figure}

\noindent\textbf{\emph{Post-Processing}}: Although MPUNet is both parameter and memory efficient, the model is trained on 2D slices with a possibly limited field of view near the boundaries. Furthermore, it is trained to segment the input viewed from different perspectives by sampling from planes of various orientations. This introduces some rotational invariance but makes it hard to distinguish between symmetric classes with very similar semantic features. For example, it is hard to be consistent with the left and right femurs when viewing the input from various perspectives. Therefore, some part of the left femur near the boundaries is mis-classified as the right femur respectively, and vice versa, as shown in Fig.~\ref{fig:post_effect}.

In order to solve this problem automatically, we propose a symmetric connected component decomposition. We only keep the largest connected component for each symmetric class pair while assigning the corresponding symmetric class value to all the other components. By doing this instead of just removing small components, those parts predicted as the left femur on the right femur are mapped correctly to the right femur, and vice versa. We then apply a standard connected component decomposition while keeping only the largest connected component for each foreground class to remove floating points (false positives). 

We acknowledge that our post-processing is highly task-specific but could also be generalized to other segmentation tasks with symmetric classes that share similar semantic features and are disconnected from each other.
\section{Data and Experiments}
We use 35 CT scans from The Cancer Imaging Archive and crop the region of interest on the images to roughly cover the area around the HJs, including the sacrum, both hip bones, and both proximal femurs~\cite{clark2013cancer}.
Each scan comprises $(415 \pm 47) \times (244 \pm 30)\times (266 \pm 29)$ voxels, with a voxel size of $(0.78 \pm 0.11)\times(0.77 \pm 0.1) \times (0.96 \pm 0.17) \text{ mm}^3$.
For the pre-training step, we use 10 scans and their associated inferior segmentations from a publicly available dataset of \textit{segmented CT pelvis scans with annotated anatomical landmarks} (CTPEL) \cite{CTPEL,wang2018pelvis_shape_iterative}.   We only use two scans with accurate segmentation to fine-tune the model in the first place, 
In the next step, two other unseen scans are used to get the segmentation results of the model. Then, we manually correct these two results and fine-tune the model again. The second fine-tuning process could be re-iterated, but four images is sufficient to obtain accurate results. We evaluate the segmentation results of our approach with minimal required fine-tuning data. A clinical expert evaluated the segmentation results of the 21 test cases.

 \noindent\textbf{\emph{Interactive Learning Setup:}}
Using prior anatomical knowledge that each class should be disconnected by at least a certain distance, contradicting cases in the model output indicate false positives (collisions) on at least one of the classes. We thus apply another Euclidean transform over the output segmentation $P$ such that each point in a predicted foreground class is mapped to the nearest distance to other foreground classes. We can then find those collision points set $E$ by applying a threshold $\epsilon$ to the distance map, as shown in Fig.~\ref{fig:comparisons} (D).
\begin{equation}
    E = \{x| P(x) \neq 0 \wedge d(x) \leq \epsilon\}
\end{equation}
Since $E$ only roughly captures the collision points, directly setting them to background will not be accurate and may introduce false negatives. 
However, the size of it ($|E|$) can be used as a metric for model performance without ground truth to decide when to terminate the interactive learning process. In our experiment where $\epsilon=2$, the model without fine-tuning gives $|E| \approx 24803$, 
while fine-tuning with two and another two accurate data reduces $|E|$ to $1000$ and $200$ respectively.

\noindent\textbf{\emph{Pre-processing}}: We pre-process the data by first filtering out all negative values in the volume because both bones and cartilages should have positive Hounsfield unit values. We then apply a standardization based on the equation $X_{\text {scale }}=(x_{i}-x_{\text {mean}})/(x_{75}-x_{25})$, where $x_{25}$ and $x_{75}$ are the 1st and 3rd quartiles respectively. This removes the median and scales the intensity based on quartiles and is more robust to outliers. No other pre-processing is applied to avoid any manual errors that can easily propagate in a neural network.

\noindent\textbf{\emph{Experimental Setup}}: The network is trained on NVIDIA GeForce RTX 3090 with a batch size of 10 using the Adam optimizer for 40 epochs with a learning rate of 1e-5 and reduced by 10\% for every two consecutive epochs without performance improvements. We apply early stopping if the performance of five consecutive epochs does not improve. Pre-training takes approximately one day, while fine-tuning takes about six hours to reach convergence.

\noindent\textbf{\emph{Augmentations}}: We follow MPUNet by applying Random Elastic Deformations to generate images with deformed strength and smoothness~\cite{simard2003best_elastic_deform} and assign a weight value of $1/3$ for the deformed samples during training~\cite{perslev2019one_multiplanarUnet}.

\section{Results}
To have suitable geometries for FE models, the auto-segmentation framework must separate bones and generate accurate results near the boundaries, which is essential for generating cartilage layers for HJ. Therefore, any standard evaluation metric such as the Dice score could be misleading. Hence, our results, including the bone outlines and the existing gap in the joints, are first validated by a senior consultant radiologist as our clinical expert.

The clinical expert initially scrolls through all the segmented slices to verify the bone contours and the gaps between the hip and femoral bones. Then, he verifies the anatomical shape and smoothness of the reconstructed 3D model. This procedure justifies  our method in obtaining precise HJ geometries.

Fig. \ref{fig:post_effect} illustrates the results of the fined-tuned model on the test set and demonstrates the effect of post-processing, where it shows that the misclassified regions in symmetric classes are successfully recovered by the post-processing step.
With the distance weight applied to loss, the model can detail the gap accurately. The final result is accurate and requires little or no human intervention for subsequent simulation experiments, e.g., FE analysis. Results in 3D are available at \href{https://github.com/MICCAI2022-155/AuToSeg}{GitHub Repo}. As an example, we have generated the cartilage geometry on the segmented HJ with a method proposed by
~\cite{moshfeghifar2022direct} to analyze the stress distributions as shown in Fig.~\ref{fig:simulation}. The results show a smooth stress pattern indicating that our method's output is suitable for use in FE simulations.

\begin{table}[hbt]
\begin{minipage}[b]{0.7\linewidth}
\centering

\begin{tabular}{lccc}
\hline
               & Dice $\uparrow$                            & GapDice $\uparrow$                         & HD($\# voxels$) $\downarrow$                           \\ \hline
Ours           & \textbf{98.63 $\pm$ 0.56} & \textbf{96.47 $\pm$ 1.60} & \textbf{3.67 $\pm$ 1.13} \\
NoPretrain     & 97.82 $\pm$ 0.59                           & 95.13 $\pm$ 1.42                           & 5.26 $\pm$ 2.10                           \\
NoWeight       & 98.12 $\pm$ 0.47                           & 94.35 $\pm$ 2.19                           & 4.58 $\pm$ 1.50                           \\
3DUNet         & 93.36 $\pm$ 1.84                           & 87.48 $\pm$ 3.01                           & 7.02 $\pm$ 1.09       \\ \hline
Ours(2)       & \textbf{97.59 $\pm$ 0.74}                       & \textbf{95.19 $\pm$ 1.14 }            & \textbf{5.18 $\pm$ 2.08}                           \\
NoPretrain(2) & 90.80 $\pm$ 9.29                           & 91.13 $\pm$ 8.53                           & 11.20 $\pm$ 7.19                          \\
NoWeight(2)   & 96.28 $\pm$ 2.91                           & 93.91 $\pm$ 1.74                           & 6.30 $\pm$ 2.95                           \\ \hline
\end{tabular}

\captionof{table}{Test Results with various design choices}
\label{tab:res}

\end{minipage}\hfill
\begin{minipage}[b]{0.28\linewidth}
\centering
\includegraphics[width=35mm]{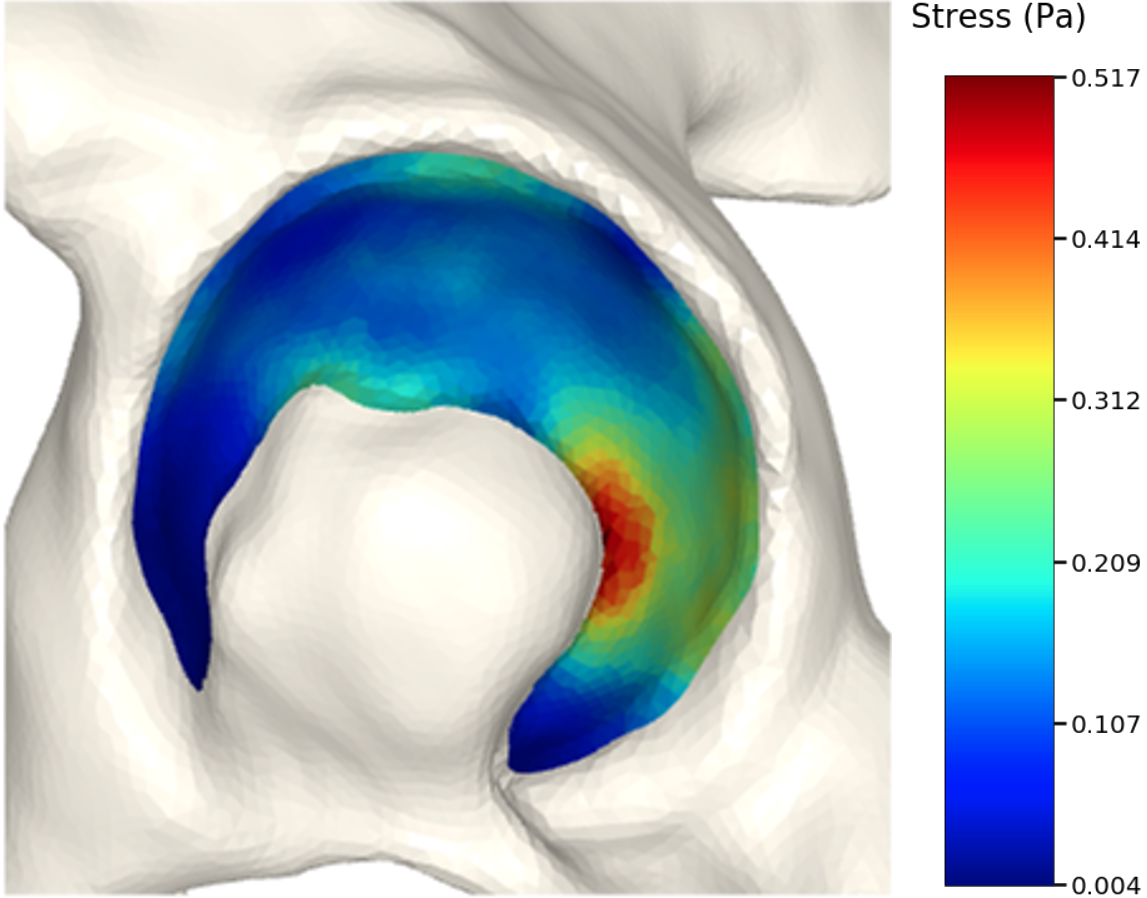}
\captionof{figure}{Smooth von Mises stress pattern}
\label{fig:simulation}
\end{minipage}
\end{table}

\subsection{Numerical Validation and Ablation Study}
Although numerical results could be misleading regarding the final FE simulations, we include them as a validation and ablation study of our several design choices. Table \ref{tab:res} shows the numerical validations on the test set, including nine images with manually corrected ground truth segmentations. We test the performance by varying one of the design choices each time while keeping the others fixed. (i) The strategy mentioned in Section \ref{sec:method} (ours), (ii) Training without using ten inaccurate public data (NoPretrain), (iii) Training without enforcing distance weight map (NoWeight), (iv) Using 3D UNet as the backbone (3DUNet). We also test and report the performance in the first stage when fine-tuned with only two manually corrected data except for (iv) because of its poor performance.

Besides the standard Dice score, we are especially interested in the surface and gap regions. Therefore, two more evaluation metrics are introduced. We use Hausdorff distance (HD) as surface measurement by computing the largest distance between the result and the nearest point on the ground truth.
\begin{equation}
    \text{HD}(P, Y) = \max(\max_{p \in P} \min_{y \in Y}\|p-y\|_{2}, \ \max_{y \in Y} \min_{p \in P}\|p-y\|_{2})
\end{equation}

We also propose a GapDice in Eq.~\eqref{eq:gapdice} to measure the average Dice score between the segmentation result and the ground truth only around the gap regions. Given the segmentation results $P$ and ground truth segmentation $Y$, we compute the Euclidean distance transformation map $Y_d$ of $Y$, corresponding to the $d_1 + d_2$ term from Eq.~\eqref{eq:weight_map}. The gap region $G$ is defined as the locations where $Y_d < \epsilon$. Dice score between $P$ and $Y$ is calculated in the standard way inside $G$. Here we choose $\epsilon=10$ as we found it to be a good indicator of both the gap and boundary regions. Fig.~\ref{fig:weight_train} (middle) shows the region computed by eroded labels, which is also an indication of $G$. 
Please refer to \href{https://github.com/MICCAI2022-155/AuToSeg}{GitHub Repo} for generated $G$.
\begin{equation}
\label{eq:gapdice}
\text{GapDice}(P, Y) = \frac{2 *|P \cap Y  \cap G|}{|P \cap G|+|Y \cap G|}
\end{equation}
The results show that MPUNet (all the first three models) works significantly better than 3D UNet in a data scarcity setting. Our pipeline outperforms in all three metrics. Especially, although the difference of the Dice score is not significant in our fine tuned model with four manually corrected data, pretraining on inaccurate data and enforcing the weight map shows a significantly better GapDice score and HD, which is vital for further simulation. The benefit of pretraining is much clearer in the first round when fine-tuned with only two accurate data, which is crucial to have minimal manual work to be fine tuned again.
 We acknowledge that the ground truth for test data is manually modified over the results from our pipeline, giving a bias when comparing multiple models, but the general goal is to show that our pipeline suits well for further simulation.

\section{Conclusion}
We  presented an auto-segmentation framework for accurate segmentation from CT scans considering the bone boundaries and inter-bone cavities. Our framework uses a modified MPUNet pre-trained on a public dataset with coarse segmentation and fine-tunes with very few data with accurate segmentation in a transfer and interactive learning setup. 
We demonstrate that our simple yet robust model can detail crucial features such as the gap where the cartilage resides.

This work is tested out on HJ CT scans and provides anatomically accurate segmentation, which has both been verified by a clinical expert and shown superior numerical results, reaching an overall Dice score above $98\%$ and above $96\%$ around gap regions. Our method can be used to enhance anatomically incorrect and poorly annotated datasets with a few accurately annotated scans. The FE analysis  shows that the generated models produce smooth stress patterns without any geometry-related artifacts. Thereby, the segmentation result of this work can be used for generating FE models with little or no manual modifications.

\subsubsection{Acknowledgements.} This project has received funding from the European Union’s Horizon 2020 research and innovation programme under the Marie Sklodowska-Curie grant agreement No. 764644. This paper only contains the author's views, and the Research Executive Agency and the Commission are not responsible for any use that may be made of the information it contains.

\bibliographystyle{plain}
\bibliography{strings}
\end{document}